\documentclass[aps,pre,reprint,twocolumn,nofootinbib]{revtex4-1}

\usepackage{subfigure}
\usepackage{amsmath}    
\usepackage{amssymb}
\usepackage{graphicx}   
\usepackage{verbatim}   
\usepackage{color}      
\usepackage{subfigure}  
\usepackage{hyperref}   

\begin{document}

\title{Ultrametricity increases the predictability of cultural dynamics}

\author{Alexandru-Ionu\c{t} B\u{a}beanu}
\affiliation{Lorentz Institute for Theoretical Physics, Leiden University, The Netherlands}
\author{Jorinde van de Vis}
\affiliation{Lorentz Institute for Theoretical Physics, Leiden University, The Netherlands}
\affiliation{Dutch National Institute for Subatomic Physics, The Netherlands}
\author{Diego Garlaschelli}
\affiliation{Lorentz Institute for Theoretical Physics, Leiden University, The Netherlands}
\affiliation{S\"aid Business School, University of Oxford, UK}
\date{\today}

\begin{abstract}
A quantitative understanding of societies requires useful combinations of empirical data and mathematical models. 
Models of cultural dynamics aim at explaining the emergence of culturally homogeneous groups through social influence. 
Traditionally, the initial cultural traits of individuals are chosen uniformly at random, the emphasis being on characterizing the model outcomes that are independent of these (`annealed') initial conditions. 
Here, motivated by an increasing interest in forecasting social behavior in the real world, we reverse the point of view and focus on the effect of specific (`quenched') initial conditions, including those obtained from real data, on the final cultural state. 
We study the predictability, rigorously defined in an information-theoretic sense, of the \emph{social content} of the final cultural groups (i.e. who ends up in which group) from the knowledge of the initial cultural traits. 
We find that, as compared to random and shuffled initial conditions, the hierarchical ultrametric-like organization of empirical cultural states significantly increases the predictability of the final social content by largely confining cultural convergence within the lower levels of the hierarchy.
Moreover, predictability correlates with the compatibility of short-term social coordination and long-term cultural diversity, a property that has been recently found to be strong and robust in empirical data. 
We also introduce a null model generating initial conditions that retain the ultrametric representation of real data. 
Using this ultrametric model, predictability is highly enhanced with respect to the random and shuffled cases, confirming the usefulness of the empirical hierarchical organization of culture for forecasting the outcome of social influence models.
\end{abstract}

\maketitle

\section{Introduction}\label{Intr}
Understanding the self-organization and emergence of large-scale patterns in real societies is one of the most fascinating, yet extremely challenging problems of modern social science \cite{thesocialatom}.
A prominent field of research studies the spontaneous emergence of groups of culturally homogeneous individuals. 
One of the mechanisms that are believed to play a key role in this process is \emph{social influence}, i.e. the gradual convergence of the cultural traits, attitudes and opinions of individuals subject to mutual social interactions. 
Stylized models of cultural dynamics under social influence have attracted the interest of an interdisciplinary community of sociologists, computational social scientists and statistical physicists \cite{Castellano}.

One of the prototypical models in this context is the popular Axelrod model \cite{Axelrod}, which has been studied in many variants over the last two decades \cite{Klemm_1, Klemm_2, Kuperman, Flache, Gonzalez-Avella, Centola, Pfau, Battiston, Stivala_2}.
The model is multi-agent, with a cultural vector associated to each agent. 
One cultural vector is a sequence of subjective cultural traits (opinions, preferences, beliefs) that each agent possesses, with respect to a predefined set of features (variables, topics, issues). 
The dynamics is driven by social influence, which iteratively increases the similarity of the cultural vectors of pairs of interacting individuals. 
However, interactions are only allowed among pairs of individuals whose vectors are already closer than a certain (implicit or explicit) threshold distance, a mechanism known as \emph{bounded confidence} and having its origins in the so-called `assimilation-contrast theory' \cite{Sherif} in social science.
The intuition behind the model, successfully confirmed via numerical simulations and analytic calculations, is that social influence increases cultural similarity, yet full convergence is precluded by bounded confidence. 
The net result is the emergence of a certain number of \emph{cultural domains}, each containing several individuals with identical cultural vectors and mutually separated by a distance larger than the bounded confidence threshold, thus no longer interacting with each other.
The value of the model is the identification of a viable, decentralized mechanism according to which cultural diversity can persist at a global (inter-domain) scale, even if it vanishes at a local (intra-domain) scale. 

Given the focus on the qualitative aspect of such an emergent pattern, the Axelrod model has been traditionally studied by specifying uniformly random initial conditions for the cultural vectors of all individuals, 
i.e. by drawing each cultural trait independently from a probability distribution that is flat over the set of possible realizations. 
Consistently with this uninformative (and deliberately unrealistic) choice, the focus of many studies has been the characterization of the outcomes of the model that are robust upon averaging over multiple realizations of the initial randomness. 
Since the cultural dynamics evolving the initial state is also stochastic, a second average over the dynamics is also required.
We may therefore say that this is the `annealed' version of the model. 
Examples of quantities that are stable across multiple realizations of uniformly random initial conditions are the expected \emph{number} and expected \emph{size} of final cultural domains. 
An obvious counter-example is the \emph{values} of the vectors ending up in such domains: as follows from the complete symmetry in cultural space implied by the uniformity of the initial randomness, 
such values are by construction maximally unpredictable.

On the other hand, recent studies have investigated the model starting from different classes of initial conditions, beyond the uniformly random one. 
In particular, emphasis has been put on using initial conditions constructed from empirical data \cite{Valori,Stivala,Babeanu_1} and their randomized, trait-shuffled counterparts 
-- obtained by randomly shuffling, for each component of the cultural vectors, the empirical values (traits) of all individuals in the sample.
These studies have emphasized a strong dependence of the final outcome on the initial conditions. 
For instance, certain model outcomes that have an interesting interpretation in terms of enabling the coexistence of short-term social collective behavior and long-term cultural diversity \cite{Valori} 
(more details are provided later in this paper) are found to vary significantly across the classes of empirical, trait-shuffled, and uniformly random initial conditions, 
while remaining largely stable when considering different instances belonging to the same class.
This stability implies that empirical cultural data share certain remarkably universal properties, 
independent of the specific sample considered and at the same time significantly different from those exhibited by random and randomized data \cite{Babeanu_1}.
This has stimulated the introduction of stochastic, structural models aimed at capturing the essential properties of the empirical cultural data \cite{Stivala, Babeanu_2}.

Strong dependence of cultural dynamics on the initial conditions might be a useful property to exploit in the light of the increasing interest towards forecasting social and cultural behavior in the real world. 
Examples include the predictability of certain aspects of political elections, public campaigns, spreading of (fake) news, financial bubbles and crashes, and commercial success of new items.
If interest is shifted towards the predictability of future long-term outcomes given certain initial conditions, then a corresponding change of perspective is implied at the level of modeling. 
In particular, the aforementioned `annealed' framework, where the outcome of models of cultural dynamics is averaged over multiple realizations of the initial randomness, becomes less relevant. 
On the contrary, if a specific (e.g. empirical) initial condition is known, it becomes natural to use it as the single initial specification of the heterogeneity of the system. 
Obviously, averaging with respect to different random trajectories of the social influence dynamics, all starting from the same initial cultural state, remains important and necessary.
We may therefore call this the `quenched' version of the model.

In this work we focus for the first time on the predictability of the \emph{social content} of the cultural domains in the final state of the Axelrod model, given a certain initial state. 
By social content we mean the composition of the different domains in terms of individuals, i.e. we are interested in forecasting `who ends up in which cultural domain'. 
It should be noted that the social content is one of those properties that, just like the values of the final cultural vectors, is maximally unpredictable when considering the usual annealed model under uniformly random initial conditions.
By contrast, we consider the quenched scenario starting from specific initial conditions sampled from empirical, shuffled, random, and an additional, `ultrametric' class of initial conditions.

We find that, remarkably, empirical and random initial conditions are associated with the highest and, respectively, lowest degree of predictability, which we rigorously define in an information-theoretic sense.
This means that, as compared with the usual uniform specification of the initial conditions of the model, empirical data allow for a much more reliable forecast of the identity of the individuals forming the final cultural domains.
We find that this result follows from the fact that the hierarchical, ultrametric-like organization of empirical cultural vectors, when coupled with bounded confidence, largely confines cultural convergence within the lower levels of the hierarchy.
This result is confirmed using surrogate data that, while retaining only the ultrametric representation of real data, are also found to be associated with a higher predictability with respect to the shuffled and random conditions.
The predictability associated to random and randomized cultural vectors is lower because it is difficult to identify a meaningful and robust hierarchical structure within the lower levels of which social influence remains confined. 

Even if we do not perform an explicit analysis of the \emph{cultural content} of the final domains, the finding that their social content is predictable, coupled with the fact that the initial cultural vectors of all individuals are known, 
implies that each final cultural vector will be a mixture of the traits of the initial vectors of the individuals ending up in the same cultural domain.
This means that, the higher the predictability of the social content, the higher that of the cultural content as well.
The take-home message is that the empirical hierarchical organization of culture and its ultrametric representation are very informative and useful for forecasting the outcome of models of cultural dynamics.

\begin{figure} 
\centering
    \includegraphics[width=9cm]{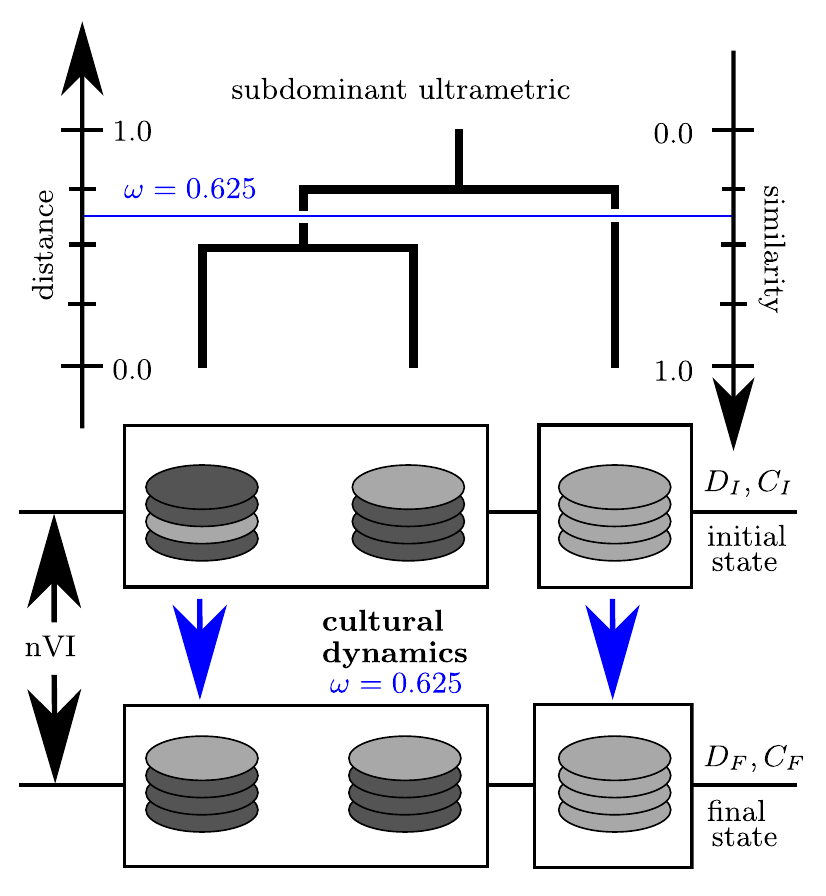} 
    \caption{Cultural dynamics with an ultrametric initial state.
At the top, a dendrogram with three leaves is shown, with a distance (or dissimilarity) scale on the left, with an associated similarity scale on the right and a threshold of $\omega = 0.625$ applied with respect to the former.
The dendrogram is a subdominant ultrametric representation of distances between three cultural vectors, which are illustrated below its branches. 
These vectors are defined in terms of four binary variables (features), corresponding to the four horizontal rows of disks, whose possible values (traits) are denoted by the light-gray and dark-gray colors. 
The boxes separate the two clusters (and connected components) obtained by applying the $\omega = 0.625$ cut in the dendrogram. 
Together, the three vectors make up an initial cultural state on which the cultural dynamics model can be applied. 
For a bounded confidence value is set to $\omega = 0.625$, one of the possible final states is shown at the bottom, 
with the boxes separating the two cultural domains.
}
    \label{Sketch} 
\end{figure}

\section{Ultrametricity and cultural dynamics}\label{UlCD}
The notion of ultrametricity refers to sets of objects that are hierarchically organized in certain abstract spaces, 
with applications in various fields, including mathematics ($p$-adic numbers), evolutionary biology (phylogenetic trees) and statistical physics (spin glasses) \cite{Rammal}.
In practice, an ultrametric representation can be produced as the output of a hierarchical clustering algorithm applied to a matrix of pairwise distances between objects \cite{Rammal}.
For the purpose of this work, these objects are the cultural vectors, and the pairwise cultural distances are computed in the same manner as in Refs. \cite{Babeanu_1, Babeanu_2, Stivala, Valori} 
-- the following explanations concerning ultrametricity are mostly restricted to cultural vectors, although many of the concepts have a wide range of applicability. 
The ultrametric representation of $N$ cultural vectors can be visualized as a dendrogram (a binary hierarchical tree; see the top of Fig. \ref{Sketch}) with $N$ leaves (one for each vector) and $N-1$ branching points 
(often referred to as ``branchings'', for simplicity), sorted by $N-1$ real numbers that are attached to them.
These numbers can be defined in two, equivalent ways: on a distance scale (top-left axis) or on a similarity scale (top-right axis) --  
both quantities take values between $0.0$ and $1.0$, while adding up to $1.0$. 
Each number is an approximation for distances between leaves that are first merged at the respective branching point.
These $N-1$ numbers and the the topology of the dendrogram retain part of the information inherent in the cultural distance matrix (which is specified by $N(N-1)/2$ numbers), so the dendrogram is an approximation of this matrix.
The approximation is exact and algorithm-independent only when the original distances are perfectly ultrametric: a stronger version of the triangle inequality is satisfied for all triplets of distinct objects \cite{Rammal}.
A cut can be performed at a certain height $\omega$ in the dendrogram, providing an $\omega$-dependent partition of the $N$ cultural vectors (see Fig. \ref{Sketch}).
For a dendrogram obtained via the single-linkage hierarchical clustering algorithm (See Ref. \cite{Sibson} and references therein), 
the $\omega$-dependent partition is the same as that encoding the connected components obtained by applying an $\omega$-threshold to the initial matrix of distances.

Ref. \cite{Valori} pointed out that a dendrogram approximating an empirical cultural state shows a clearer hierarchical organization than those approximating its shuffled or random counterparts, 
suggesting that the ultrametric representation is better suited for empirical data than for shuffled or random data. 
In addition, cultural dynamics applied to the empirical cultural state appeared to mostly induce convergence within the clusters of the $\omega$-dependent partition, 
if $\omega$ is equal to the bounded confidence threshold used in the cultural dynamics model (see below),
These observations were made in a qualitative way, by visually inspecting dendrograms obtained with the average-linkage hierarchical clustering algorithm \cite{Anderberg, Sokal}.
Instead, we perform here a systematic, quantitative comparison between $\omega$-dependent partitions of initial cultural states and associated partitions of final states resulting from cultural dynamics,
for different classes of initial cultural states.
In addition, one of these classes is defined by enforcing, on average, the ultrametric representation of empirical data, generalizing a method originally proposed in Ref. \cite{Tumminello} for biological taxonomies. 
Whenever an ultrametric representation is constructed within this study, the single-linkage algorithm \cite{Sibson} is used instead of the average-linkage one, 
since it provides the subdominant ultrametric, which is the `closest from below' to the original distances and unique \cite{Rammal_2}, 
while also being equivalent to the hierarchical connected-component representation, as mentioned above. 
This choice is also common for the purpose of evaluating measures of ultrametricity, like the cophenetic correlation coefficient, 
which is done in Ref. \cite{Stivala}.

\begin{figure*}
\centering
	\subfigure[][]{\includegraphics[width=8.5cm]{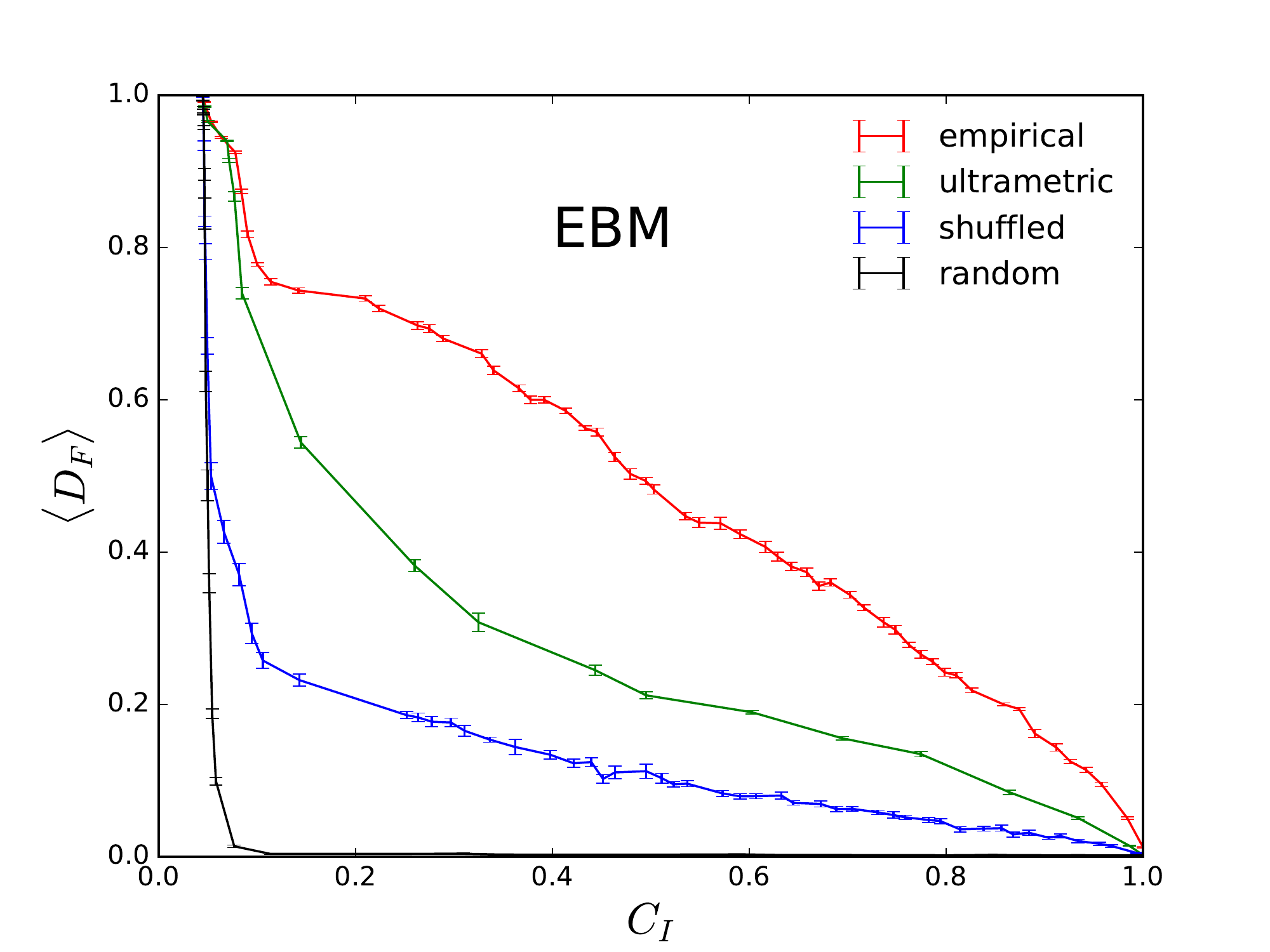}\label{DvC}}
	\subfigure[][]{\includegraphics[width=8.5cm]{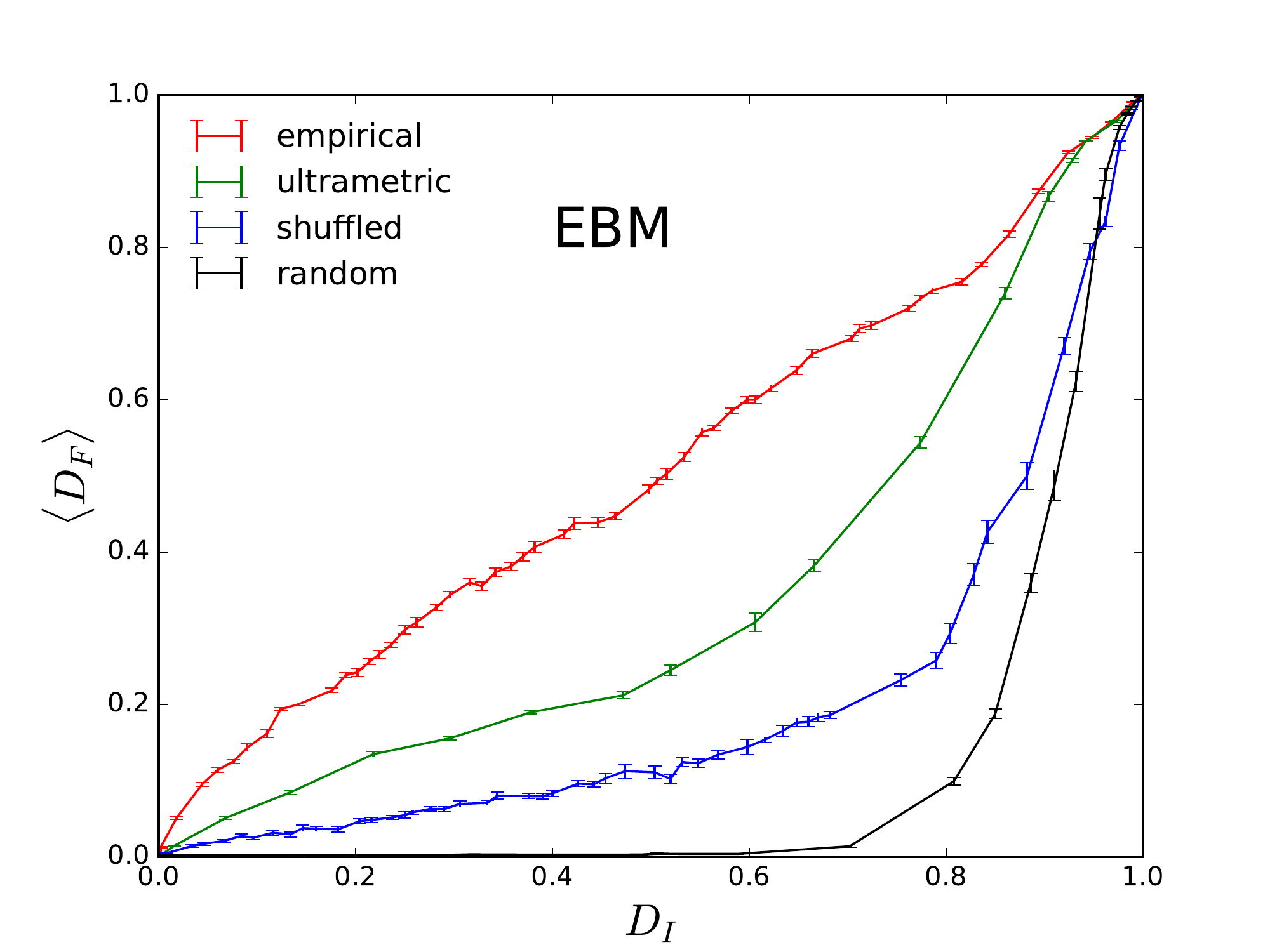}\label{DvNC}} 
	\caption{Relationships between the important diversity and coordination measures.
One sees the dependence of the final, average diversity $\langle D_F \rangle$, 
first \subref{DvC} on the initial coordination $C_I$, 
second \subref{DvNC} on the initial diversity measure $D_I$.
This is shown for one empirical (red), one ultrametric-generated (green), one shuffled (blue) and one random (black) set of cultural vectors.
All sets of cultural vectors have $N=500$ elements and are defined with respect to the same cultural space, from the variables of the empirical Eurobarometer (EBM) data. 
The errors of $\langle D_F \rangle$ are standard mean errors obtained from 10 cultural dynamics runs. 
}
\label{DvCvNC}
\end{figure*} 

Cultural dynamics is modeled here by a simple, Axelrod-type model, without any underlying geometry for a social network or a geographical-physical space:
essentially, all $N$ agents are connected to each other. 
Instead, a bounded-confidence threshold $\omega$ is present, controlling the maximum cultural distance for which social influence can successfully occur. 
This is exactly the model used in Refs. \cite{Valori, Babeanu_1, Babeanu_2} and partly in Ref. \cite{Stivala}.
As anticipated in Sec. \ref{Intr}, this model converges to a random final, absorbing state, one that consists of domains of internally identical and mutually non-interacting cultural vectors -- 
distances within such groups are zero, while distances across are larger or equal to $\omega$.

Fig. \ref{Sketch} captures the essence of this study.
At the center, the figure shows an initial cultural state with 3 vectors, defined in terms of 4 binary features, with possible traits (values) denoted by the two shades of gray.
Each of the three vectors is matched to a branch of the dendrogram drawn at the top, which encodes the subdominant ultrametric representation of the initial cultural state.
For this specific case, the distance between the first two vectors is $0.5$, while the distances between any of these two and the third are $0.75$, 
which together make up a perfectly ultrametric discrete space, thus exactly matching the distances encoded by the dendrogram.
The horizontal line denotes a possible $\omega$-cut that can be applied to the dendrogram, which induces a splitting into two (in the example shown) branches and two associated subsets of vectors, 
which together form a $\omega$-dependent partition (or clustering) of the initial set.
This partition is the same as that induced by the set of connected cultural components of the $\omega$-thresholded cultural graph. 
At the bottom, the figure shows one possible final state resulting from the cultural dynamics process, 
for a bounded confidence threshold set to the same $\omega$ value as the dendrogram cut. 
The groups of identical vectors constitute another, $\omega$-dependent partition characterizing the cultural state, which exactly matches, in this case, the initial state partition.
Other final configurations are possible, due to the stochastic nature of cultural dynamics.
It is even possible, although unlikely, that by a succession of convenient interactions the second vector ``migrates'' from the cluster on the left to the one on the right during the dynamics. 
The abundance of such deviations is quantitatively studied below, for several classes of initial conditions.

\section{Partition-specific quantities}\label{PSQ}
The initial and final partitions form the basis of all calculations performed in this study.
Each type of partition is characterized by two types of quantities, denoted by $(D_I, C_I)$ for initial partitions and by $(D_F, C_F)$ for final partitions.
These quantities are referred to as the coordination ($C_I$ and $C_F$) and the diversity measures ($D_I$ and $D_F$).
They are computed according to the following formulas:
\begin{equation}	
  \label{DaC}
	D_a(\omega) = \frac{N_C^a(\omega)}{N}, \quad C_a(\omega) = \sqrt{\sum_A \left(\frac{S_A^a}{N}\right)^2_{\omega}},
\end{equation}
where $a \in \{I,F\}$	distinguishes between ``initial'' and ``final'', $N_C^a$ is the number of clusters (connected components if $a=\mathrm{I}$, groups of identical vectors if $a=\mathrm{F}$),
and $S_A^a$ is the size of cluster $A$ for the given $\omega$ value. 
Note that $D_a$ is a measure of diversification, while $C_a$ is a measure of non-homogeneity encoded by the respective partition.
Moreover, since cultural dynamics is a stochastic process, it is meaningful to talk about averages over final state partitions (over multiple dynamical runs),
which is particularly useful for the final diversity measure $\langle D_F(\omega) \rangle$.

The $\langle D_F(\omega) \rangle$ quantity has been interpreted as a measure of propensity to long-term cultural diversity, while the $C_I(\omega)$ has been interpreted as a measure of propensity to short-term collective behavior \cite{Valori, Babeanu_1}. 
Through their common dependence on $\omega$, the correspondence between the two quantities is graphically illustrated in Fig. \ref{DvC}.
Along each curve, different points correspond to different $\omega$ values, while different curves correspond to different classes of initial conditions.
It is clear that the empirical cultural state allows for much more compatibility between the aspects measured by the two quantities than the shuffled and the random cultural state, as pointed out in Ref. \cite{Valori}.
In fact, this is the analysis used in Ref. \cite{Valori} to highlight the structure of empirical cultural data and in Ref. \cite{Babeanu_1} to emphasize the universality of this structure -- 
except for the ``ultrametric'' scenario, which is first introduced here. 
In this scenario, a set of $N$ cultural vectors is generated such that, on average, 
the pairwise distances reproduce those encoded in the subdominant ultrametric representation of an empirical set of cultural vectors of the same $N$.
This is  achieved using an extension of the method developed in Ref. \cite{Tumminello}, in the context of genetic sequences. 
The extension here allows the method to work with combinations of features of different ranges and types, 
where the range stands for the number of traits and the type indicates whether the feature is ordinal or nominal.  
This is described in detail in Appendix \ref{AppSUG}.
On the other hand, a shuffled set of cultural vectors is obtained by randomly and independently permuting empirical cultural traits among vectors, with respect to every feature, thus exactly enforcing the empirical trait frequencies.
Note that the ultrametric cultural state comes closer to the empirical behavior than the shuffled cultural state, 
suggesting that empirical ultrametric is better than empirical trait frequencies at explaining the generic empirical structure.
Finally, a random set of cultural vectors is obtained by drawing each trait at random, from a uniform probability distribution, while only retaining the empirical data format, 
or cultural space -- the number of features, together with the range and type of each feature, thus the. 
Eurobarometer 38.1 \cite{EBM} data is used here, formatted according to the procedure in Ref. \cite{Babeanu_1}

For the same four sets of cultural vectors used in Fig. \ref{DvC}, 
the average final diversity $\langle D_F(\omega) \rangle$ is plotted against the initial diversity $D_I(\omega)$ in Fig. \ref{DvNC}.
This visualization, previously used \cite{Valori,Stivala} without the ultrametric scenario, illustrates the extent to which cultural dynamics preserves the number of clusters when going from the initial to the final partition.
As observed before, the number of clusters is well preserved by cultural dynamics acting on empirical data, which happens much less for shuffled data and even less for random data.
This goes along with the idea that the final partition can be predicted from the initial partition if empirical data is used for specifying the latter.
Note that, like in Fig. \ref{DvC}, ultrametric-generated data lies in between the empirical and shuffled scenarios, confirming that the subdominant ultrametric information, which is directly related to the sequence of $\omega$-dependent initial partitions, is rather robust with respect to cultural dynamics.

\begin{figure}
\centering
	\includegraphics[width=9cm]{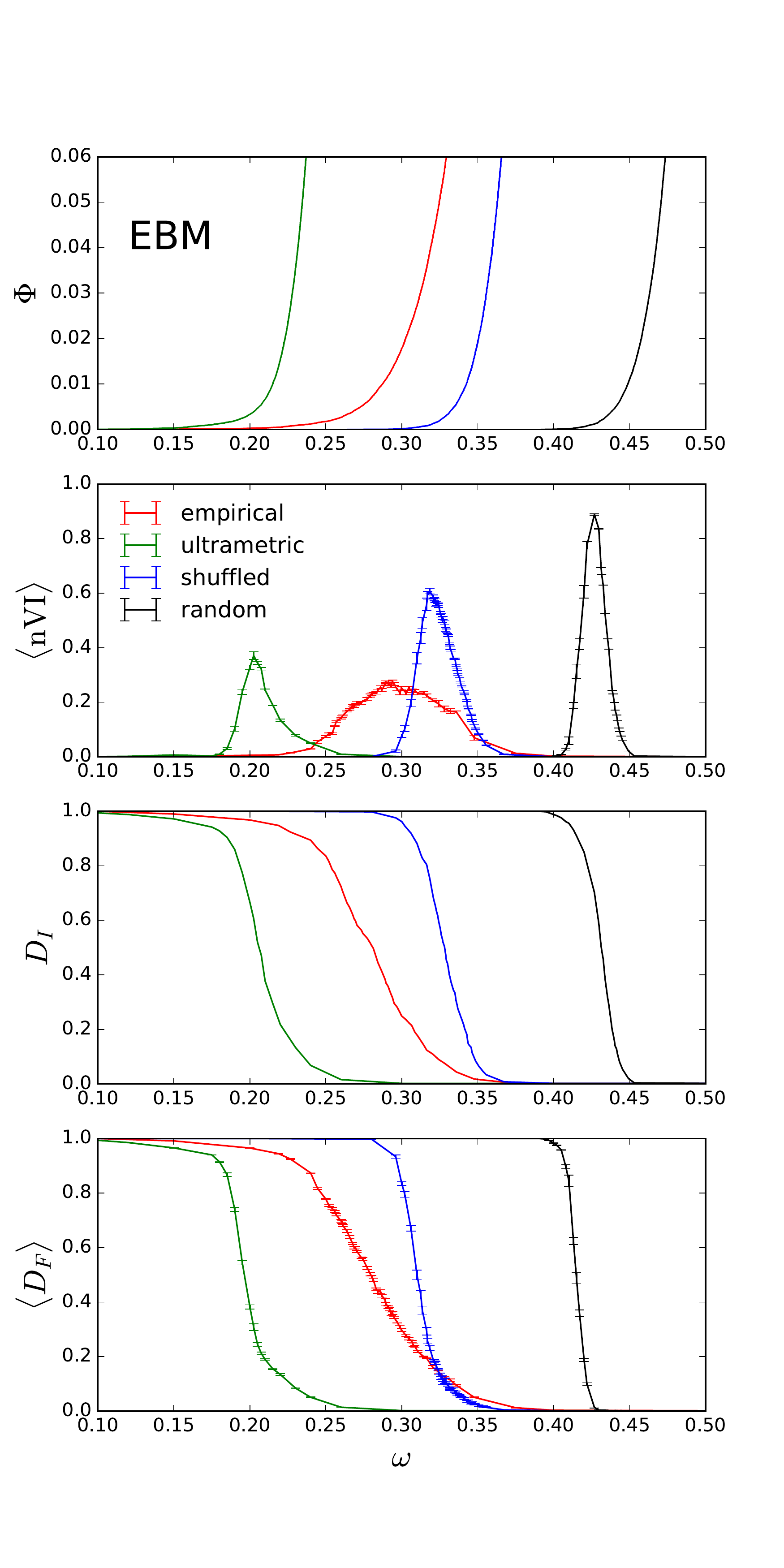}
	\caption{Visualization of the ultrametric predictability of cultural dynamics.
The dependence on the bounded-confidence threshold $\omega$ is shown for several quantities:
most importantly, the normalized variation of information between the initial and final partitions $\langle \mathrm{nVI} \rangle$ at the center-top;
the fraction of initially active cultural links $\Phi$ at the top;
the initial diversity $D_I$ at the center-bottom;
the final, average diversity $\langle D_F \rangle$ at the bottom.
This is shown for one empirical (red), one ultrametric-generated (green), one shuffled (blue) and one random (black) set of cultural vectors.
All sets of cultural vectors have $N=500$ elements and are defined with respect to the same cultural space, from the variables of the Eurobarometer (EBM) data. 
The errors of $\langle D_F \rangle$ and $\langle \mathrm{nVI} \rangle$ are standard mean errors obtained from 10 cultural dynamics runs. 
}
\label{ThrDep}
\end{figure}

\section{Predictability of the final state}
Although informative, the comparison between the $\langle D_F(\omega) \rangle$ and $D_I(\omega)$ is incomplete as a way of assessing the predictability of the final partition from the initial partition: 
two partitions might have the same number of clusters, but the sizes and/or contents of these clusters might be very different. 
In order to take all this into account in a consistent way, the discrepancy between the initial and final state partitions is evaluated using the variation of information measure $\mathrm{VI}$ \cite{Meila_2}, as a function of$\omega$.
This is an information-theoretic measure that acts as a metric distance within the space of possible partitions of a set of $N$ elements. 
It is convenient to work with the normalized version of this quantity $\mathrm{nVI}(\omega) = \mathrm{VI}(\omega)/\log(N)$, 
which retains the meaning and metricity of the original quantity, as long as $N$ remains the same ($N=500$ for all results presented here). 

The dependence of $\langle\mathrm{nVI}\rangle$ on $\omega$ is shown in the second panel of Fig. \ref{ThrDep},
for the same 4 cultural states used in Fig. \ref{DvCvNC}, where the averaging is performed over multiple dynamical runs, like for the $\langle D_F \rangle$ quantity.
The empirical state shows the lowest maximal $\langle\mathrm{nVI}\rangle$ value, followed by the ultrametric, the shuffled and the random states.  
This figure shows, in a rigorous way, that the outcome of cultural dynamics can be predicted relatively well based on the initial state, if this is constructed from empirical data and comparably well if this is constructed based on the empirical ultrametric information. 
On the other hand, shuffled and random data exhibit lower predictability. 
Note that, for either scenario, $\langle\mathrm{nVI}\rangle$ vanishes for the low-$\omega$ and the high-$\omega$ regions, 
which is where both the initial and final partitions consist of $N$ single-object clusters and of one, $N$-objects cluster respectively.
This can be understood by looking at the dependence of the $D_I$ and $\langle D_F \rangle$ quantities on $\omega$ shown in the in the third and fourth panels:
the $\omega$ region for which $\langle \mathrm{nVI} \rangle$ is significantly larger than $0.0$ is roughly the region where either $D_I$ or $\langle D_F \rangle$ is substantially different from $1.0$ and $0.0$.

In parallel, the first panel of Fig. \ref{ThrDep} shows the $\omega$-dependence of the fraction of initially active cultural links $\Phi$: the fraction of pairs $(i,j)$ of cultural vectors whose distance $d_{ij} < \omega$ in the initial state.
This shows that the $\omega$ interval that is non-trivial with respect to $D_I$, $\langle D_F \rangle$ and $\langle\mathrm{nVI}\rangle$ seems to be largely determined by the shape of $\Phi$, which is nothing else than the cumulative distribution of intervector distances.
The properties of this distribution -- 
average lower for empirical data than for random data, standard deviation higher for empirical data than for either shuffled or random data --
have been studied before \cite{Valori, Stivala} and are recognizable in the first panel of Fig. \ref{ThrDep}.
Note that, for the ultrametric scenario, the interesting $\omega$ region and the $\Phi$ profile are compressed in a lower-$\omega$ region compared to empirical data.
This means that the branchings in the dendrogram obtained from ultrametric-generated data occur at lower $\omega$ values than those in the dendrogram obtained from the original, empirical data.
In turn, this is due to the distances between the ultrametric-generated cultural vectors reproducing, on average, the subdominant ultrametric empirical distances, rather than the original empirical distances,
while the former are known to systematically underestimate the latter, particularly for higher distance values, as long as the empirical vectors are not perfectly ultrametric, 
which in practice is always the case.

There is another aspect that can be noted when comparing, for either scenario, the shape of $\Phi(\omega)$ in the first panel
with the shape of $D_I(\omega)$ in the third panel of Fig. \ref{ThrDep}:
as $\omega$ is decreased, most of the cultural links need to be eliminated in order to reach the abrupt region of the $D_I(\omega)$ transition,
for which the number of clusters in the initial partition becomes comparable to $N$.
This is not surprising on general grounds. 
For instance, the Erd\H{o}s-R\'{e}niy model of random graphs \cite{Erdos} exhibits a critical link density of $1/N$, at which a giant connected component is present,
if $N$ is the number of nodes in the graph, instead of the number of cultural vectors. 
Still, this analogy should not be taken too far.
The random graph interpretation is closest to the random cultural state scenario used here,
since the expected pairwise distance entailed by the latter is the same for any pair of cultural vectors, 
just like the connection probability entailed by the former is the same for any pair of nodes. 
However, even the random scenario has a the metric structure, due to how cultural spaces are defined\cite{Babeanu_1},
which should introduce more triangles than expected otherwise,
while the shuffled and empirical scenarios are additionally affected by inhomogeneities in their cultural space distributions.

\begin{figure}[h]
\centering
	\includegraphics[width=9cm]{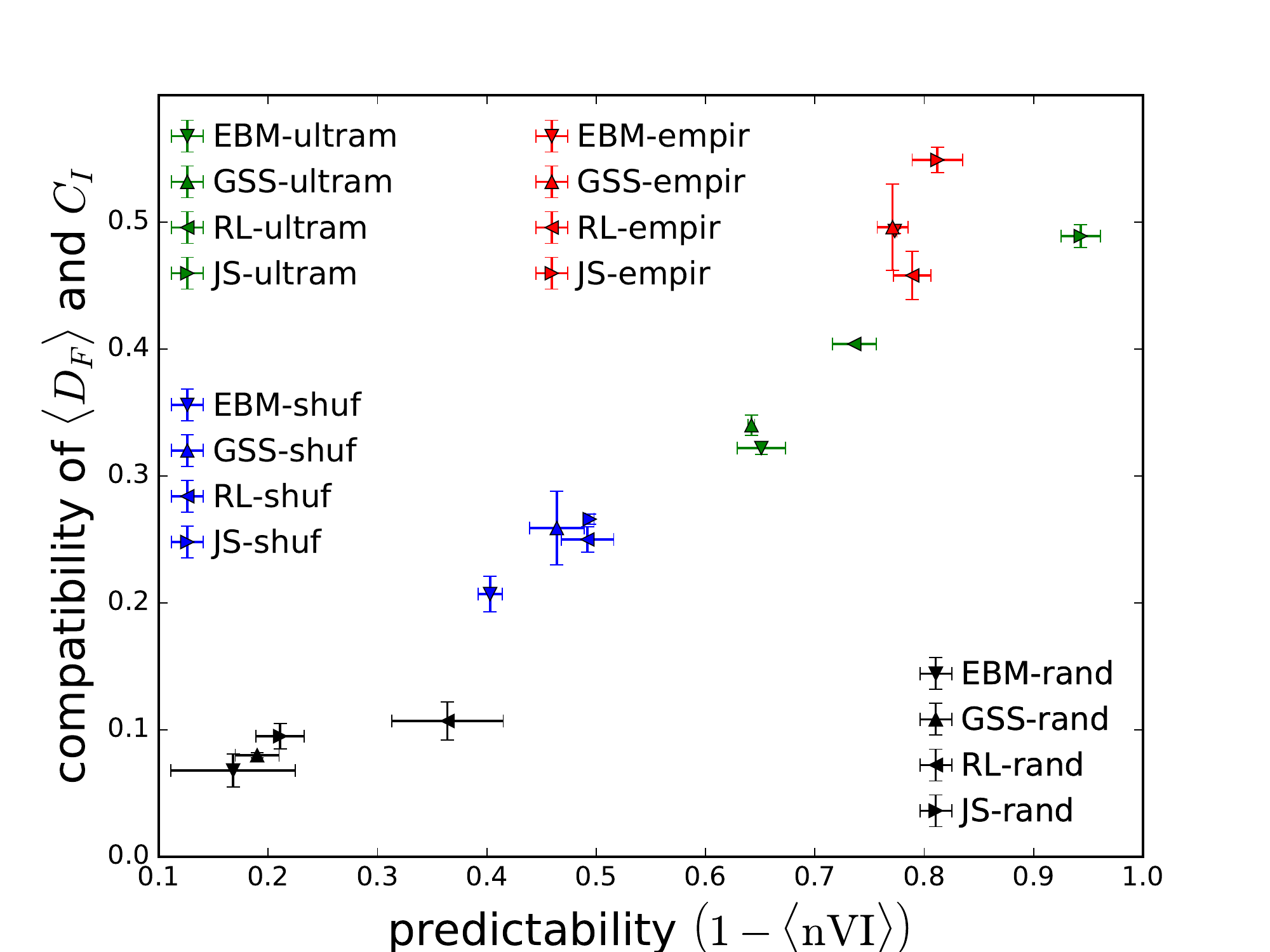}
	\caption{ Relationship between compatibility of final diversity and initial coordination (vertical axis) and predictability of the final partition from the initial partition. 
Each point corresponds to one cultural state, belonging to one class and to one empirical source: each color corresponds to one class of cultural states, while marker type correspond to one dataset, as indicated in the legends. 
All cultural state consist of $N=500$ cultural vectors.
}
\label{VoI-DvC}
\end{figure} 

The analysis presented in Figs. \ref{DvCvNC} and Fig. \ref{ThrDep} was repeated for three other datasets: the General Social Survey \cite{GSS}, Jester\cite{JS} and the Religious Landscape \cite{RL}, 
processed according to the formatting rules of Ref. \cite{Babeanu_1}. 
For all four datasets, the results are presented in a joint, compact manner by means of Fig. \ref{VoI-DvC}, while more detailed results are shown in Appendix \ref{AppDR}.
Each of the points in the figure corresponds to a combination of one dataset and one scenario.
The vertical axis corresponds to a measure of compatibility between long-term cultural diversity $\langle D_F \rangle$ and short-term collective behavior $C_I$,
namely a measure of the overall departure of the $\langle D_F \rangle \, \mathrm{vs} \ C_I$ curve from the lower-left corner in Fig. \ref{DvC}.
The horizontal axis corresponds to a measure of predictability of the final state from the initial state,
namely an inverse measure of the overall departure of the $\langle\mathrm{nVI}\rangle \, \mathrm{vs} \ \omega$ from the horizontal axis in the second panel of Fig. \ref{ThrDep}. 

For both measures, simple definitions are employed: rather than integrating information from every $\omega$ value for which some departure is present,
both definitions conceptually rely only on one, representative $\omega^*$ point, for which both departures are relatively high. 
Specifically, $\omega^*$ is defined by intersecting the $\langle D_F \rangle \, \mathrm{vs} \ C_I$ curve with the main diagonal $\langle D_F \rangle = C_I$.
In practice, since just a finite number of $\omega$ values are available for any combination of dataset and scenario, 
one uses instead the two $\omega$ values that are closest to the main diagonal of the $\langle D_F \rangle \, \mathrm{vs} \ C_I$ plot from either of the two sides.
These two values, labeled as $\omega_L$ and $\omega_R$, ``bracket'' $\omega^*$ from the left and right respectively: $\omega_L < \omega^* < \omega_R$.
The $\omega^*$ itself is never explicitly calculated, but is conceptually useful for the explanations below. 

The compatibility approximates the distance between the $(\langle D_F(\omega^*) \rangle \, \mathrm{vs} \ C_I(\omega^*))$ point and the $(\langle D_F \rangle = 0, C_I = 0)$ point,
normalized by the length of the main diagonal of the $\langle D_F \rangle \, \mathrm{vs} \ C_I$ plot.
In practice, this is evaluated in terms of $\omega_L$ and $\omega_R$ according to:
\begin{equation}
 \frac{\sqrt{\langle D_F (\omega_L) \rangle^2 + C_I^2(\omega_L)} + \sqrt{\langle D_F (\omega_R) \rangle^2 + C_I^2(\omega_R)}}{2\sqrt{2}}, \nonumber
\end{equation}
while the associated error is evaluated as:
\begin{equation}
 \frac{\sqrt{\langle D_F (\omega_L) \rangle^2 + C_I^2(\omega_L)} - \sqrt{\langle D_F (\omega_R) \rangle^2 + C_I^2(\omega_R)}}{2\sqrt{2}}. \nonumber
\end{equation}
The predictability approximates the distance between the $(\omega^*, \langle\mathrm{nVI}(\omega^*)\rangle)$ point and the $\langle\mathrm{nVI}\rangle = 1$ line.
In practice, this is evaluated as: 
\begin{equation}
	1-\frac{\langle\mathrm{nVI}(\omega_L)\rangle + \langle\mathrm{nVI}(\omega_R)\rangle}{2}, \nonumber
\end{equation}
while the associated error is evaluated as: 
\begin{equation}
	\frac{\lvert \langle\mathrm{nVI}(\omega_L)\rangle - \langle\mathrm{nVI}(\omega_R)\rangle \lvert}{2}. \nonumber
\end{equation}

Note that compatibility increases with predictability in a roughly linear way,
at least for the cultural states considered here.
Moreover, cultural states belonging to the same class tend to cluster together in the compatibility-predictability space.
A notable exception is ultrametric-Jester, 
which is significantly outside the ultrametric class in terms of predictability, showing higher predictability than any of the empirical states. 
Still, it is clear that cultural states that are closer to the universal $\langle D_F \rangle \, \mathrm{vs} \ C_I$ empirical behavior 
also allow for better estimates of the final partition from the initial one.

The observed increase of compatibility with predictability provides some insights about the nature of empirical data, 
or at least about the shape of an empirical-like dendrogram characteristic for the upper-right corner of Fig. \ref{VoI-DvC}. 
This can be understood by realizing that the ultrametric and empirical states approach an ideal, limiting situation of perfect predictability, for which the initial and final partitions are identical irrespective of $\omega$.
This implies that $\langle D_F(\omega) \rangle = D_I(\omega)$ and consequently that the $\langle D_F \rangle \, \mathrm{vs} \ C_I$ curve is essentially the $D_I \, \mathrm{vs} \ C_I$ curve and thus controlled by the geometry of the subdominant ultrametric dendrogram.
One can then show -- see Appendix Sec. \ref{AppDT} -- that this geometry needs to be highly ``unbalanced'' in order to explain the close-to-linear $\langle D_F\rangle \approx 1 - C_I$ empirical behavior in Fig. \ref{DvC} and the compatibility values of approximately $0.5$ following from it.
For a perfectly-unbalanced geometry, the $k$th highest dendrogram branching separates only one leaf from the remaining $N-k$, for all $k \in \{1,...,N-1\}$.
By contrast, a perfectly-balanced geometry entails a splitting into two, equal clusters for each dendrogram branching, 
which would induce an inverse square $\langle D_F \rangle \propto C_I^{-2}$ behavior -- see Appendix Sec. \ref{AppDT} -- closer to that of shuffled and random cultural states, with a lower compatibility value.
Thus, while going from the random to the empirical class, by enforcing more and better empirical information, the increasing level of compatibility becomes more suggestive of an unbalanced dendrogram geometry, while the increasing level of predictability increases the reliability of this geometric interpretation.

\section{Conclusion}
This study focused on the ultrametric representation of sets of cultural vectors used for specifying the initial state of cultural dynamics models.
On one hand, it introduced another procedure for randomly generating initial conditions based on the subdominant ultrametric information of empirical data.
On the other hand, it examined the extent to which the subdominant ultrametric representation can be used for predicting the final state of cultural dynamics in a simple theoretical setting. 
The bounded-confidence threshold parameterising the dynamical model was used to extract an initial-state partition from the ultrametric representation.
This was sistematically compared, in terms of variation if information, with the corresponding final state partition consisting of groups of identical cultural vectors.
The comparison showed that the predictive power of the ultrametric is relatively high for empirical cultural states, which are closely followed by ultrametric-generated states, which are followed by the shuffled and then by the random states. 
Moreover, higher predictability appears to go hand in hand with higher compatibility between a propensity to long-term cultural diversity and a propensity to short-term collective behaviour,
which was previously shown to be a hallmark of empirical structure.
This means that ultrametric information is better than trait-frequency information at explaining this structure.
These results further advance the understanding of the relationship between ultrametricity and cultural dynamics. 
Moreover, it is tempting to speculate that, for the purpose of forecasting the dynamics of culture in the real world, 
knowledge about the current distribution of individuals in cultural space might be sufficient, with little or no need for running simulations, 
at least if one assumes that consensus-favoring social influence is the essential driving force of this dynamics.  

{\bf Acknowledgements:} 

AIB acknowledges discussions Leandros Talman.
DG acknowledges financial support from the Dutch Econophysics Foundation (Stichting Econophysics, Leiden, the  Netherlands).
This work was also supported by the Netherlands Organization for Scientific Research (NWO/OCW).

\bibliographystyle{unsrt}
\bibliography{Paper}{}



\appendix

\section{Ultrametric-generation method}
\label{AppSUG}

This section explains the method for generating sets of cultural vectors belonging to the ``ultrametric'' class.
The method is an extension of that developed in Ref. \cite{Tumminello}.
The description here is somewhat similar to that in Ref. \cite{Tumminello}, but the nomenclature specific to cultural vectors is used, instead of that specific to genetic sequences.

The method takes as input a dendrogram, as well as a target cultural space -- the number of cultural features $F$, together with the range (number of traits) $q$ and type (nominal or ordinal) of each feature. 
This information is taken from empirical data and the single-linkage hierarchical clustering alorithm is employed for constructing the dendrogram whenever the method is used in this study. 
Upon every use, the method generates, in a stochastic way, a set of $N$ cultural vectors associated to the $N$ leaves of the dendrogram, such that, on average,
the pairwise similarities between cultural vectors match the similarities encoded by the dendrogram. 

More precisely, for each cultural feature in the target space, the method enforces: 
\begin{equation}
	\label{purpose}
	E[s^q_{ij}] = \rho_{\alpha_{ij}},
\end{equation}
where $E[...]$ stands for ``expectation value'', $\alpha_{ij}$ is the lowest branching in the dendrogram joining leaves $i$ and $j$, 
$\rho_{\alpha_{ij}}$ is the similarity encoded by this branching
and $s^q_{ij}$ is the partial contribution to the similarity between cultural vectors $i$ and $j$ of a feature of range $q$, 
which is computed according to the following formula:
\begin{equation}
	\label{feat_sim}
	s^q_{ij} = 
	\begin{cases} 
		\delta(x_i^k, x_j^k)	& \text{if nominal}, \\
		\frac{|x_i^k - x_j^k|}{q^k-1} 	& \text{if ordinal},
	\end{cases}
\end{equation}
which depends on whether the feature is nominal or ordinal.
Eq. \eqref{feat_sim} is consistent with the cultural distance definition in Refs. \cite{Valori, Stivala, Babeanu_1, Babeanu_2} 
(as mentined above: $\mathrm{similarity} = 1.0 - \mathrm{distance}$).

In Eq. \eqref{purpose}, the expectation $E[...]$ implies averaging over multiple runs of the method, for the same dendrogram and the same cultural feature. 
Although in practice the method is used only once (and independently) for each feature, the fact that a large number $F$ of features are present makes this approach sensible:
the expectation $E[s_{ij}]$ of the complete similarity $s_{ij}$ will also match $\rho_{\alpha_{ij}}$
(since the complete similarity is the arithmetic average of the feature-level similarities),
while the fluctuations of $s_{ij}$ around $\rho_{\alpha_{ij}}$ with $F$.
In other words, as pointed out in Ref. \cite{Tumminello}, the expectation in Eq. \eqref{purpose} can be interpreted in two idealized ways: averaging over infinitely many runs or averaging over infinitely many features.

In order to enforce Eq. \eqref{purpose} for every pair $(i,j)$, the method controls for the extent to which the traits of different vectors are choosen independently of each other. 
For every feature, all the $N$ choosen cultural traits originate in independent random draws from a uniform probability distribution, but the number of draws is smaller or equal to $N$.
Thus, the traits of vectors $i$ and $j$ either originate in the same draw, with probability $P_{ij}$, 
or originate in different draws, with probability $1-P_{ij}$.
In the former case the two traits are identical, with a well-determined feature-level similarity $s^q_{ij} = 1$.
In the latter case, the two traits may be identical or different, so that $s^q_{ij}$ fluctuates around an expectation value $f(q)$.
Taking both cases into account, the expectation value of $s^q_{ij}$ is:
\begin{equation}
	\label{expect}
	E[s^q_{ij}] = P_{ij} + [1 - P_{ij}] f(q),
\end{equation}
where the expectation for different draws $f(q)$ reads:
\begin{equation}
	\label{func_f}
	f(q) = 
	\begin{cases} 
		\frac{1}{q} 			& \text{if nominal}, \\
		\frac{2q - 1}{3q} 		& \text{if ordinal},
	\end{cases}
\end{equation}
which is the expression of the expected, feature-level similarity between two traits drawn at random from a uniform probability distribution,
obtained analytically from Eq. \eqref{feat_sim}.
The choices of traits and the associated random draws are mangaged by the stochastic-algorithmic part of the method (briefly explained at the end of this section), which is designed to ensure that:
\begin{equation}
	\label{prop}
	P_{ij} = \rho_{\alpha_{ij}}^I 
\end{equation}
is satisfied, where $\rho_{\alpha_{ij}}^I$ is a corrected version of the similarity $\rho_{\alpha_{ij}}$ implicit in the ${\alpha_{ij}}$ branching:
\begin{equation}
	\label{height}
	\rho_{\alpha_{ij}}^I = \rho_{\alpha_{ij}} - h(\rho_{\alpha_{ij}},q),
\end{equation}
where $h$ is a correction function chosen such that Eqs.~\eqref{purpose} holds, subject to \eqref{expect} and \eqref{prop}. 
Specifically, by combining Eq.~\eqref{prop} with Eq.~\eqref{expect} and then with Eq.~\eqref{purpose}, one obtains:
\begin{equation}
	\label{relation}
	\rho_{\alpha_{ij}}^I + [1 - \rho_{\alpha_{ij}}^I] f(q) = \rho_{\alpha_{ij}}.
\end{equation}		
By inserting Eq.~\eqref{height} in Eq.~\eqref{relation} and further manipulations, one obtains the following expression for the correction function: 
\begin{equation}
	\label{func_h}
	h(\rho_{\alpha},q) = \frac{1 - \rho_{\alpha}}{1 - f(q)} f(q).
\end{equation}
Note that Eq.~\eqref{prop} identifies $\rho_{\alpha_{ij}}^I$ with a probability, 
meaning that $\rho_{\alpha}^I > 0$ should be satisfied for all branchings $\alpha$.
This implies, given Eq.~\eqref{height} and Eq.~\eqref{func_h}, 
that $\rho_{\alpha} > f(q)$ for all branchings $\alpha$ of the given dendrogram and for all features in the target space.
This condition needs to be satisfied in order for this method to be valid and is actually satisfied by all four empirical dendrograms used in this study.
Also note that the method in Ref.~\cite{Tumminello} is recovered as a special case of the above, 
by restricting to nominal features of constant $q$ via Eq.~\eqref{func_f}. 

Finally, it is worth describing the stochastic-algorithmic part of the method.
For each of the $F$ features in the target space, the following steps are carried out:
\begin{itemize}
	\item the dendrogram is recursively explored starting with the root branching; 
	for every branching $\alpha$ reached by this exploration, one of the following two things happens: 
	\begin{itemize}
		\item one of the $q$ traits is randomly chosen, according to a uniform distribution and assigned to all cultural vectors corresponding to leaves under branching $\alpha$, without further exploring any branching below $\alpha$;
		\item the exploration is continued with each of the two branches emerging from $\alpha$, if that branch leads to another branching, instead of leading to a leaf;
	\end{itemize}
	with probability $Q_{\alpha}$ for the former and probability $1-Q_{\alpha}$ for the latter, where:
	\begin{equation}
		Q_{\alpha} = \frac{\rho_{\alpha}^I - \rho_{p(\alpha)}^I}{1 - \rho_{p(\alpha)}^I},
	\end{equation}
	where $p(\alpha)$ is the parent branching of $\alpha$, if $\alpha$ is not the root, while $\rho_{p(\alpha)}^I = 0$ if $\alpha$ is the root.  
	\item for each of the leaves whose traits are not assigned during the above step, one of the $q$ traits is randomly chosen, according to a uniform distribution and assigned to the respective cultural vector.
\end{itemize}
This algorithmic procedure ensures that Eq.~\ref{prop} holds, for reasons that are fully explained in Ref.~\cite{Tumminello}.
		
It is worth noting that the ultrametric-generation method described in this section makes use of all the information inherent in the geometry of the dendrogram that it receives as input 
-- both the topology and the similarities $\rho$ encoded by the branching points of the dendrograms are used.
However, the generated sets of cultural vectors will in general not be precisely ultrametric, in the strict mathematical sense \cite{Rammal}
(unless it is applied in the limit of $F$ being much larger than $N$).
Still, they are generated based on the empirical ultrametric information and are arguably as close as they can be to reproducing the ultrametric set of pairwise distances.

\section{Detailed results}
\label{AppDR}

\begin{figure}
\centering
	\includegraphics[width=9cm]{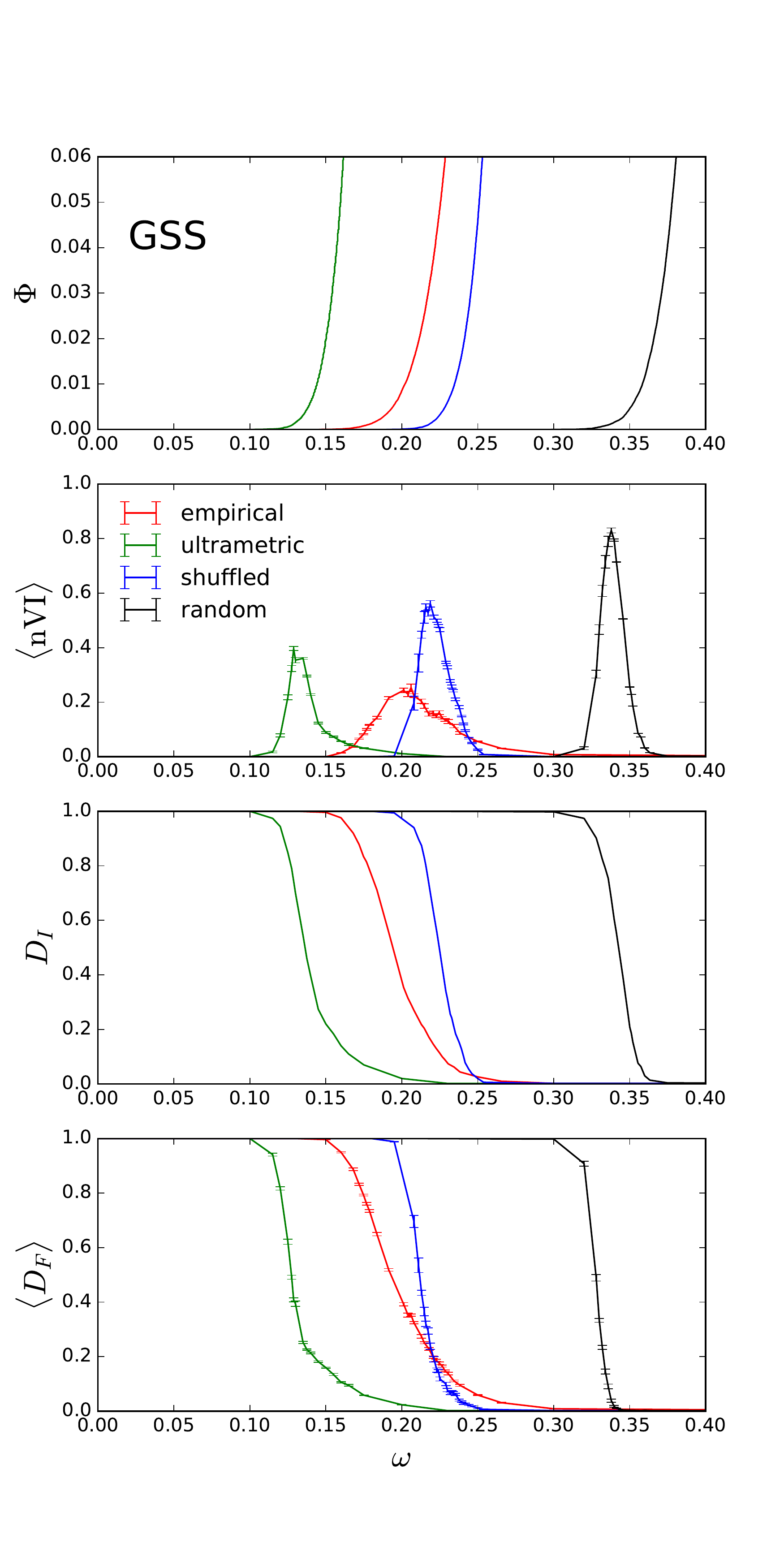}
	\caption{Vizualisation of the ultrametric predictability of cultural dynamics.
The dependence on the bounded-confidence threshold $\omega$ is shown for several quantities:
most importantly, the normalized variation of information between the initial and final partitions $\langle\mathrm{nVI}\rangle$ at the center-top;
the fraction of initially active cultural links $\Phi$ at the top;
the initial diversity $D_I$ at the center-bottom;
the final, average diversity $\langle D_F \rangle$ at the bottom.
This is shown for one empirical (red), one ultrametric-generated (green), one shuffled (blue) and one random (black) set of cultural vectors.
All sets of cultural vectors have $N=500$ elements and are defined with respect to the same cultural space, from the variables of the General Social Survey (GSS) data. 
The errors of $\langle D_F \rangle$ and $\langle\mathrm{nVI}\rangle$ are standand mean errors obtained from 10 cultural dynamics runs. 
}
\label{AppThrDep_1}
\end{figure}

\begin{figure}
\centering
	\includegraphics[width=9cm]{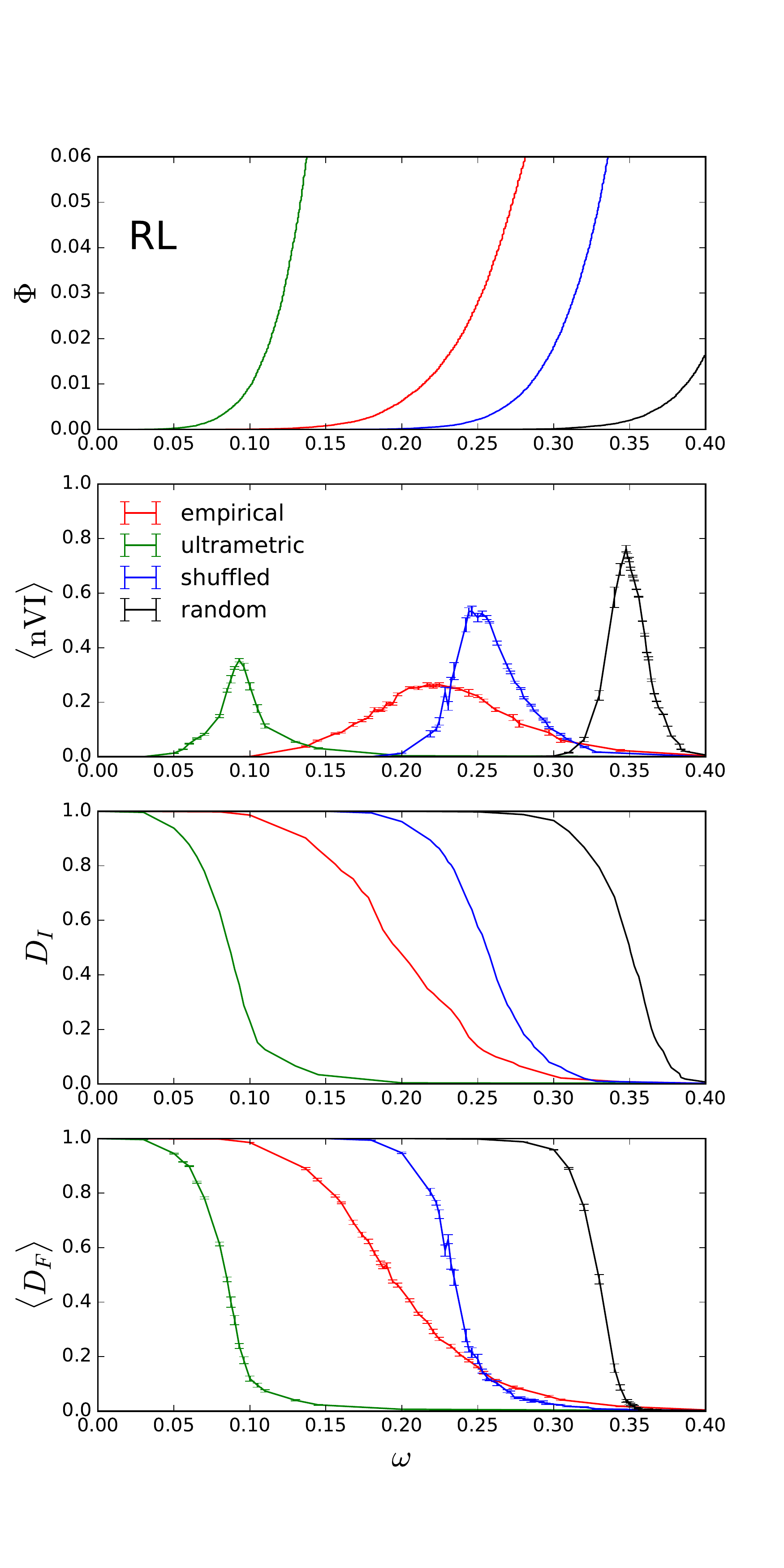}
	\caption{Vizualisation of the ultrametric predictability of cultural dynamics.
The dependence on the bounded-confidence threshold $\omega$ is shown for several quantities:
most importantly, the normalized variation of information between the initial and final partitions $\langle \mathrm{nVI} \rangle$ at the center-top;
the fraction of initially active cultural links $\Phi$ at the top;
the initial diversity $D_I$ at the center-bottom;
the final, average diversity $\langle D_F \rangle$ at the bottom.
This is shown for one empirical (red), one ultrametric-generated (green), one shuffled (blue) and one random (black) set of cultural vectors.
All sets of cultural vectors have $N=500$ elements and are defined with respect to the same cultural space, from the variables of the Religious Landscape (RL) data. 
The errors of $\langle D_F \rangle$ and $\langle \mathrm{nVI} \rangle$ are standand mean errors obtained from 10 cultural dynamics runs. 
}
\label{AppThrDep_2}
\end{figure}

\begin{figure}
\centering
	\includegraphics[width=9cm]{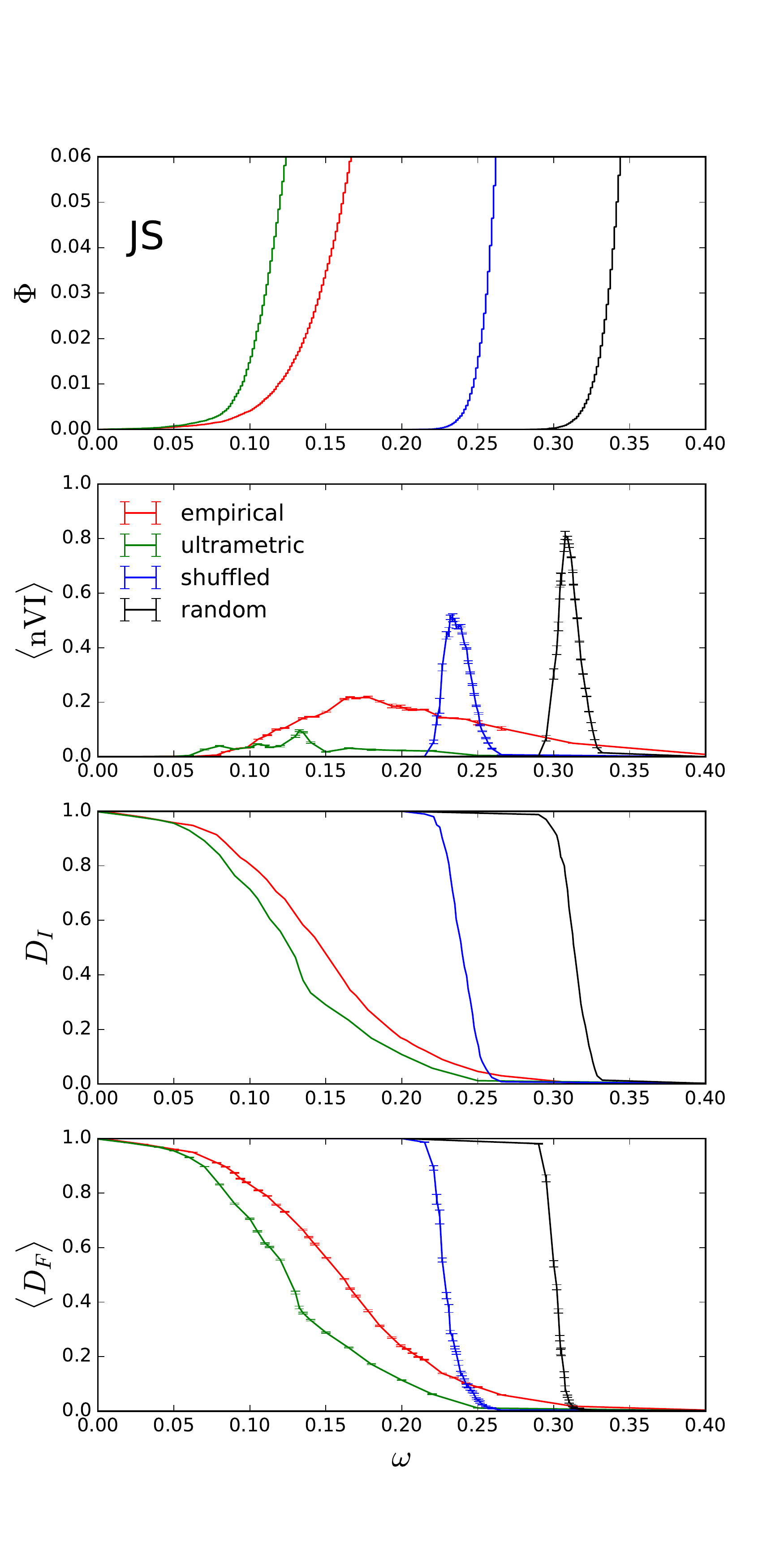}
	\caption{Vizualisation of the ultrametric predictability of cultural dynamics.
The dependence on the bounded-confidence threshold $\omega$ is shown for several quantities:
most importantly, the normalized variation of information between the initial and final partitions $\langle \mathrm{nVI} \rangle$ at the center-top;
the fraction of initially active cultural links $\Phi$ at the top;
the initial diversity $D_I$ at the center-bottom;
the final, average diversity $\langle D_F \rangle$ at the bottom.
This is shown for one empirical (red), one ultrametric-generated (green), one shuffled (blue) and one random (black) set of cultural vectors.
All sets of cultural vectors have $N=500$ elements and are defined with respect to the same cultural space, from the variables of the Jester (JS) data. 
The errors of $\langle D_F \rangle$ and $\langle \mathrm{nVI} \rangle$ are standand mean errors obtained from 10 cultural dynamics runs. 
}
\label{AppThrDep_3}
\end{figure}

This section shows the complete results concerning the $\omega$-dependence of relevant quantities, 
for the other three data sets that are used in this study in addition to the Eurobarometer (EBM \cite{EBM}): 
the General Social Survey (GSS \cite{GSS}) data in Fig. \ref{AppThrDep_1}, the Religious Landscape (RL \cite{RL}) data in Fig. \ref{AppThrDep_2} and the Jester (JS \cite{JS}) data in Fig. \ref{AppThrDep_3}. 
Each of these three figures follows the format of Fig. \ref{ThrDep} above, with four panels and four scenarios.
Although, for each type of scenario, there is a certain variability in the width and location of the non-trivial $\omega$ interval, 
the results are qualitatively similar to those obtained for EBM data,
with a notable exception visible for the analysis of Jester data in Fig. \ref{AppThrDep_3}:
the second panel shows that the discrepancy between the initial and the final partition, as measured by $\langle\mathrm{nVI}\rangle$,
is clearly smaller for the ultrametric cultural state than for the empirical cultural state, so the overal predictability is higher. 
This is in agreement with the observation made in relation to Fig. \ref{VoI-DvC} about the relatively high predictability value of the Jester-ultrametric point.

\section{Dendrogram geometry}
\label{AppDT}

This section gives some analytical insight on how the dendrogram geometry is related to the behaviour of the two measures of initial diversity $D_I$ and initial coordination $C_I$.
As functions of $\omega$, the two measures only change (in steps) when $\omega$ crosses the distance value associated to any of the branchings of the dendrogram. 
Thus, one can replace the dependence of $D_I$ and $C_I$ on $\omega$ with a dependence on $k$, which counts the number of dendrogram branchings above a given $\omega$, in terms of their associated distance values --
$k$ increases from $0$ to $N-1$ as $\omega$ decreases from $1.0$ to $0.0$. 
Based on Eq.~\eqref{DaC}, one can thus write:
\begin{equation}
\label{discDC}
	D_I(k) = \frac{N_C^I(k)}{N}, \quad C_I(k) = \sqrt{\sum_A \left(\frac{S_A^I}{N}\right)^2_{k}}.
\end{equation}

\begin{figure} 
\centering
    \includegraphics[width=9cm]{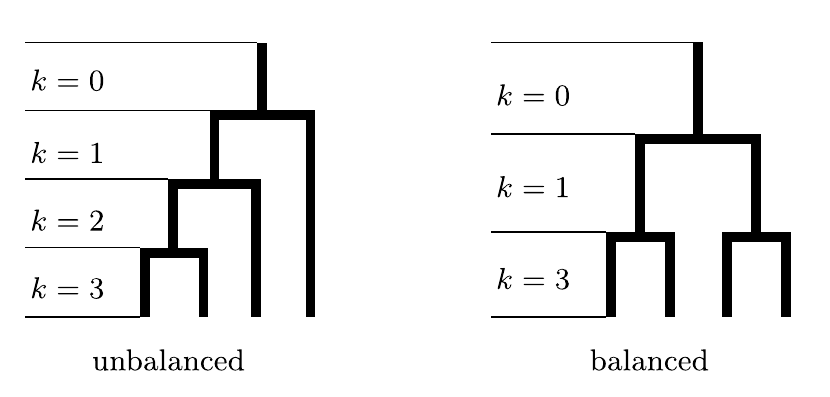} 
    \caption{Sketch of a ``perfectly balanced'' (left) dendrogram geometry and a ``perfectly unbalanced'' (right) one, 
for $N=4$ leaves. The values of $k$ indicate the number of branchings above any cut that would be applied to the dendrogram within the respective horizontal band.
}
    \label{Sketch_2} 
\end{figure}

There are two, extreme types of dendrogram geometries that are worth considering, the "perfectly-unbalanced geometry" and the "perfectly-balanced geometry".
These are illustrated in Fig. \ref{Sketch_2}. 

For the perfectly-unbalanced geometry, shown on the left side of Fig. \ref{Sketch_2}, the number of connected components is:
\begin{equation}
\label{discNC}
	N_C^I(k) = k+1,
\end{equation}
while the sizes of the connected component are: 
\begin{equation}
\label{discSA}
	S_A^I(k) =     
		\begin{cases} 
      N-k, 	& \mathrm{if } A = 1  \\ 
      1, 	& \mathrm{if } A \in\{2,3,...,k+1\}
    \end{cases}.
\end{equation}
From Eqs.~\eqref{discDC} and \eqref{discNC}, one obtains the behaviour of the initial diversity measure:
\begin{equation}
\label{ubalD}
	D_I(k) = \frac{k+1}{N},
\end{equation}
while from Eqs.~\eqref{discDC} and \eqref{discSA} one obtains the behaviour of the initial coordination measure:
\begin{equation}
	C_I(k) = \sqrt{\left(\frac{N-k}{N}\right)^2 + k\left(\frac{1}{N}\right)^2},
\end{equation}
from which it follows that:
\begin{equation}
	C_I(k) = \sqrt{1 - 2\frac{k}{N} + \frac{k^2}{N^2} + \frac{k}{N^2}},
\end{equation}
where one can neglect the $\frac{k}{N^2}$ term in the limit of large $N$, thus obtaining: 
\begin{equation}
\label{ubalC}
	C_I(k) \approx 1 - \frac{k}{N}.
\end{equation}
From Eqs. \ref{ubalD} and \ref{ubalC} it follows that:
\begin{equation}
	C_I(k) \approx 1 - D_I(k) - \frac{1}{N},
\end{equation}
which can be rephrased, after neglecting the $\frac{1}{N}$ term in the limit of large $N$, to:
\begin{equation}
	D_I(k) \approx 1 - C_I(k),
\end{equation}
which describes the second-diagonal empirical behaviour of Fig. \ref{DvC}, under the assumption that $D_F(k) = D_I(k), \forall k$.

For a perectly-balanced geometry, shown on the right side of Fig. \ref{Sketch_2}, the only relevant values of $k$ (those corresponding to non-vanishing $\omega$ intervals) are $k = \sum_{i=0}^{l-1}2^i$, with $l \in \{0,1,2,...,\log_2{N}\}$.
For these values of $k$, the number of connected components, like in the unbalanced case, is described by Eqs.~\eqref{discNC},
while the sizes of the connected components are:
\begin{equation}
\label{discSA2}
	S_A^I(k) = N/(k+1), \forall A \in \{1,2,...,k+1\},
\end{equation}
from which it follows that the initial coordination measure is:
\begin{equation}
	\label{balC}
	C_I(k) = \sqrt{(k+1)\left(\frac{1}{k+1}\right)^2} = \frac{1}{\sqrt{k+1}}.
\end{equation}
Since the $k$-dependence of the initial diversity measure $D_I$, like in the unbalanced case, is described by Eq. \eqref{ubalD}, it follows that:
\begin{equation}
  D_I(k) = \frac{1}{N C_I^2(k)},
\end{equation}
which, under the assumption that $D_F(k) = D_I(k), \forall k$, entails a curve more similar to that of the shuffled or random curves of Fig.~\ref{DvC}, than to that of the empirical curve. 
Moreover, this curve comes arbitrarily close to the lower left corner as $N$ increases. 

To sum up, the above reasoning shows that, as long as $D_F(\omega) = D_I(\omega), \forall \omega$,
an unbalanced dendrogram geometry fits the empirical $D_F(C_I)$ behaviour very well, while a balanced dendrogram geometry does not.
Although the latter entails a $D_F \propto C_I^{-2}$ behaviour quite similar to that observed for shuffled or random data, one cannot say that a balanced geometry is a good description for either of these two cases,
since the assumption that $D_F = D_I$ is false for both these cases, for the interesting $\omega$ intervals.

\end{document}